\documentclass[sigconf]{acmart}
\usepackage[utf8]{inputenc} % allow utf-8 input
\usepackage[T1]{fontenc}    % use 8-bit T1 fonts
\usepackage{url}            % simple URL typesetting
\usepackage{booktabs}       % professional-quality tables
\usepackage{amsfonts}       % blackboard math symbols
\usepackage{nicefrac}       % compact symbols for 1/2, etc.
\usepackage{microtype}      % microtypography

\usepackage{amsthm}
\usepackage{amsmath}
\usepackage{cleveref}

\usepackage[figuresright]{rotating}
\usepackage{caption}
\usepackage{subcaption}
\usepackage{multirow}
\usepackage[ruled,vlined]{algorithm2e}

\usepackage{xcolor}         % colors
\usepackage{float}

\usepackage{graphics}
\usepackage{subcaption}
\usepackage{multirow}
\usepackage{ctable}

\usepackage{xcolor}
\usepackage{textcomp}
\usepackage{multicol}

\definecolor{blueannoback}{RGB}{234,242,250}
\definecolor{greenannoback}{RGB}{230,244,214}
\definecolor{redannoback}{RGB}{255, 230, 230}
\definecolor{yellowannoback}{RGB}{252, 251, 202}
\definecolor{orangeannoback}{RGB}{247, 189, 114}
\definecolor{pinkannoback}{RGB}{238, 172, 252}

\usepackage{listings}

\lstdefinelanguage{prompt}{
    morecomment=[l][\textcolor{BurntOrange}]{@}
}

\lstdefinestyle{prompt-style}{
    language=prompt,
    escapeinside={\%*}{*)}
}

\newcommand{\vgap}{\vspace{-0.3cm}}
\AtBeginDocument{%
  }

\setcopyright{acmcopyright}
\begin{document}

\title{Towards Quantum Range Query for Spatial-Temporal-Semantic Trajectory Data}

\author{Hao Li}
\affiliation{%
  \institution{National University of Singapore}
  %\city{Madison}
  \country{Singapore}
  }
%\email{hao.li@nus.edu.sg}

\author{Zhihang Liu}
\affiliation{%
  \institution{The Chinese University of Hong Kong}
  %\city{Madison}
  \country{Hong Kong}
}
%\email{zhihangliu@link.cuhk.edu.hk}

\author{Liwei Zou}
\affiliation{%
  \institution{The Hong Kong University of Science and Technology (Guangzhou)}
  %\city{Madison}
  \country{China}
  }
%\email{lzou260@connect.hkust-gz.edu.cn} 

\author{Jinlin Wu}
\affiliation{%
  \institution{The Hong Kong University of Science and Technology (Guangzhou)}
  \country{China}
  }
%\email{jwu923@connect.hkust-gz.edu.cn}

\author{Wufan Zhao}
\affiliation{%
  \institution{The Hong Kong University of Science and Technology (Guangzhou)}
  \country{China}
  }
%\email{wufanzhao@hkust-gz.edu.cn}

%%
%% The "author" command and its associated commands are used to define
%% the authors and their affiliations.
%% Of note is the shared affiliation of the first two authors, and the
%% "authornote" and "authornotemark" commands
%% used to denote shared contribution to the research.
% \author{Ben Trovato}
% \authornote{Both authors contributed equally to this research.}
% \email{trovato@corporation.com}
% \orcid{1234-5678-9012}
% \author{G.K.M. Tobin}
% \authornotemark[1]
% \email{webmaster@marysville-ohio.com}
% \affiliation{%
%   \institution{Institute for Clarity in Documentation}
%   \streetaddress{P.O. Box 1212}
%   \city{Dublin}
%   \state{Ohio}
%   \country{USA}
%   \postcode{43017-6221}
% }

%%
%% By default, the full list of authors will be used in the page
%% headers. Often, this list is too long, and will overlap
%% other information printed in the page headers. This command allows
%% the author to define a more concise list
%% of authors' names for this purpose.
\renewcommand{\shortauthors}{Li et al.}

%%
%% The abstract is a short summary of the work to be presented in the
%% article.
\begin{abstract}

Range query is a fundamental task in geospatial data search and many other downstream applications. Classic range queries often rely on tree-based spatial indexes, of which the query speed depends on the number of indexed points $k$ within the queried range. For instance, a classical B+ tree answers a range query in $O(\log N+k)$. For a long time, this speed has long been considered asymptotically optimal in classic database systems, until the recent emergence of quantum computing, where a quantum B+ tree may requires only  $O(\log_B N)$. This paper presents Quantum Range Query (QRQ) via a hybrid quantum-classic algorithm to return the range query results in \textit{quantum superpositions}. In this context, QRQ is designed to accelerate classic range query on spatial-temporal-semantic trajectory geodata using quantum algorithms. Specifically, QRQ develops quantum variants of R-tree, TB-tree and KD-tree, where the physical slots of a node, including unused padding slots, are treated as an array that a quantum random-access memory (QRAM) can read in superpositions. Evaluations on three common trajectory datasets, namely GeoLife, T-Drive, and GDP Drifter, and \(10{,}000\) queries per setting, the QRQ speedup at \(1\%\) target selectivity ranges from \(2.03\times\) to \(64.70\times\). More importantly, QRQ shows a great potential in optimizing existing trajectory range query, so it becomes the shared primitive on which human mobility analysis, and later searches specified by large language models (LLMs) and geo-foundation models (GeoFMs), can rest.

\end{abstract}

%%
%% The code below is generated by the tool at http://dl.acm.org/ccs.cfm.
%% Please copy and paste the code instead of the example below.
%%

\vspace{-0.2cm}
\begin{CCSXML}
<ccs2012>
   <concept>
       <concept_id>10010147.10010257.10010293.10010294</concept_id>
       <concept_desc>Computing methodologies~Neural networks</concept_desc>
       <concept_significance>500</concept_significance>
       </concept>
   <concept>
       <concept_id>10010147.10010257.10010293.10010319</concept_id>
       <concept_desc>Computing methodologies~Learning latent representations</concept_desc>
       <concept_significance>500</concept_significance>
       </concept>
   <concept>
       <concept_id>10010405.10010432.10010437</concept_id>
       <concept_desc>Applied computing~Earth and atmospheric sciences</concept_desc>
       <concept_significance>500</concept_significance>
       </concept>
 </ccs2012>
\end{CCSXML}

\ccsdesc[500]{Computing methodologies~Neural networks}
\ccsdesc[500]{Computing methodologies~Learning latent representations}

%%
%% Keywords. The author(s) should pick words that accurately describe
%% the work being presented. Separate the keywords with commas.
% \keywords{Large language model, geospatial artificial intelligence, few-shot learning, task-agnostic training}
\keywords{Quantum Algorithm, Range Query, Spatio-temporal Data, GeoAI }
%% A "teaser" image appears between the author and affiliation
%% information and the body of the document, and typically spans the
%% page.
% \begin{teaserfigure}
%   \includegraphics[width=\textwidth]{sampleteaser}
%   \caption{Seattle Mariners at Spring Training, 2010.}
%   \Description{Enjoying the baseball game from the third-base
%   seats. Ichiro Suzuki preparing to bat.}
%   \label{fig:teaser}
% \end{teaserfigure}

%%
%% This command processes the author and affiliation and title
%% information and builds the first part of the formatted document.
\maketitle
\raggedbottom

\vspace{-0.2cm}

\section{Introduction} \label{sec:intro}
Spatiotemporal trajectories are now a primary observational record of cities and environmental systems~\cite{gonzalez2008mobility,song2010predictability}, where each trajectory is a polyline of samples \(p=(x,y,t)\). For trajectory geodata, a range query asks a precise question that given a spatiotemporal window, which segments intersect this window? The required answer is the exact set of intersecting segments, rather than an approximate or nearest-neighbor match. Range queries are fundamental operations, where mobility analysis, forecasts of congestion, and reconstructions of environmental transport all rest on~\cite{zheng2015trajectory}. As the archives grow, the efficiency of the range query, rather than the analysis that follows it, becomes a bottleneck of the speed of mobility analysis.

\begin{figure*}[t]
  \centering
  \includegraphics[width=0.8\textwidth]{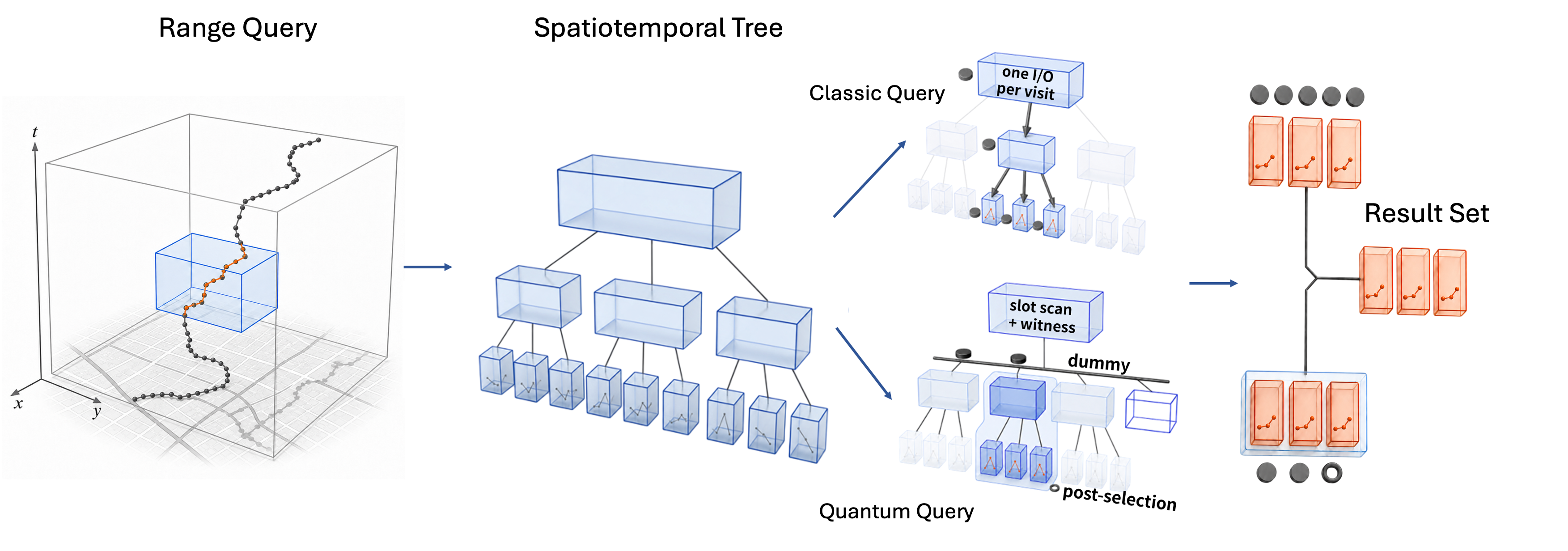}
  \vgap
  \caption{QRQ indexes 3-D trajectory segments with R-tree, TB-tree, and KD-tree. Herein, classical range query will cost one I/O per visited node, while the quantum model, scans physical slots (including padding slots) and incurs a post-selection cost after a containment or intersection witness. Both procedures share the same answer set \(\mathcal{A}(Q)\).}
  \Description{Method overview: a space-time query cube, a shared 3-D MBR tree labeled Packed R, TB, and KD, a classical pointer walk, a quantum slot scan with a contained witness, and identical orange hit segments.}
  \label{fig:method}
\end{figure*}

For decades, spatiotemporal range query relies on tree-based indexes which address the range query task with a pruned walk over spatiotemporal partitions. For example, packed R-trees and TB-trees prune with three-dimensional minimum bounding rectangles, so that a child node is opened only when its box meets the query \(Q\)~\cite{guttman1984r,pfoser2000novel}. In this way, a range query becomes efficient as internal nodes reject a large fraction of the archive. The residual cost is overlap among those rectangles. The R$^*$-tree already treats that overlap as a structural cost of spatial partitioning~\cite{beckmann1990rstar}, where a box may intersect \(Q\) while most of its interior lies outside \(Q\), every such child must still be opened, and the $k$ hits must still be enumerated at the leaves. To this reason, the speed of classical range query, e.g., $O(logN+k)$, therefore depends the visited volume ($k$) rather than with tree height alone. The index guarantees the exact set, but it does not eliminate the listing cost.

Fortunately, recent quantum algorithms show promising potential in reducing this listing cost by interrogating unstructured memory into superpositions, so that a query need not inspect every candidate such as in a classic range query~\cite{montanaro2016quantum}. Herein, an ideal quantum random-access memory (QRAM) attaches that capability for addressable storage~\cite{giovannetti2008quantum}. In this context, Quantum-based trees (e.g., B+ tree \citep{liu2024quantumbptree}) can avoid a full listing of the \(k\) candidates, using a slot-wise scan and a post-selection over superpositions. Specifically, quantum algorithms can extract marked candidates after the first containment (or so-called intersection witness), but they assume one-dimensional keys whose routing intervals are disjoint during the range query~\cite{liu2024quantumbptree}.

Inspired by this observation, this paper presents a hybrid Quantum Range Query (QRQ) method aiming at accelerating spatio-temporal-semantic range queries over large-scale trajectory geodata. Specifically, QRQ implements quantum versions of tree-based range query algorithms, including R-tree, TB-tree and KD-tree, where the physical slots of a node, including unused padding slots, are treated as an array that a QRAM can read in its superpositions. Tested on three trajectory dataset, namely GeoLife, T-Drive, and GDP Drifter, preliminary experiment results are reported using \(10{,}000\) queries per datasets, the speedup of three QRQ variants ranges from \(2.03\times\) to \(64.70\times\), confirming the capability of QRQ to accelerate this fundamental task of spatiotemporal data range queries for numerous downstream applications \citep{kazemi2012geocrowd}. Moreover, we envision that QRQ will also benefit those more sophisticated search functions, including intensive semantic search with large language models (LLMs) and geospatial foundation models (GeoFMs), given its improved query speed.

% Methodology section for ``Towards Quantum Range Query for Spatial-Temporal-Semantic Trajectory GeoData''
% This section assumes that algorithm2e, amsmath, amssymb and hyperref are loaded.

\section{Methodology}
\label{sec:methodology}

\subsection{Problem Formulation}

Define a trajectory record as $r_i=(\tau_i,\mathbf{p}_i,\mathbf{s}_i,\mathrm{id}_i)$, comprising temporal ($\tau_i$), spatial ($\mathbf{p}_i$), and semantic ($\mathbf{s}_i$) components. The index key is $z_i=\phi(r_i)\in\mathbb{Z}^{d}$, where $\phi(\cdot)$ is either a space-filling embedding or a quantized attribute concatenation. For a query window $\mathcal{W}=[\boldsymbol{\ell},\boldsymbol{u}]$, the classical answer set is:
\begin{equation}
  \mathcal{A}(\mathcal{W})= \{(z_i,r_i):\ell_j\leq z_{i,j}\leq u_j,\ j=1,\ldots,d\}.
\end{equation}

The target quantum answer is the normalized superposition:
\begin{equation}
  |\Psi_{\mathcal{W}}\rangle =\frac{1}{\sqrt{k}} \sum_{(z_i,r_i)\in\mathcal{A}(\mathcal{W})}|z_i\rangle|r_i\rangle.
\end{equation}

This state serves as an intermediate input for downstream quantum operations (e.g., machine learning or temporal aggregation) rather than for direct measurement. If classical materialization is required, the classical listing cost $\Omega(k)$ is unavoidable.

To optimize state preparation, we partition the index into a classical control plane and a quantum data plane. The classical plane leverages conventional routing bounds and node occupancy metadata to isolate a compact candidate set. The quantum plane stores corresponding node relations and payloads via QRAM. This separation is critical as it exploits classical index geometry to prune the search space before conducting quantum parallelism, avoiding inefficient unstructured searches across all $N$ records.

\subsection{Quantum Index Representation}

Assuming a node capacity of $B$ and identifiers in $\{0,\ldots,M-1\}$, let $C(v,j)$ and $P(v,j)$ denote the $j$-th child and leaf entry of node $v$, respectively. We define two QRAM oracles:
\begin{equation}
  \mathcal{Q}_{0}|vB+j\rangle|0\rangle = |vB+j\rangle|C(v,j)\rangle,
\end{equation}
\begin{equation}
  \mathcal{Q}_{1}|vB+j\rangle|0\rangle = |vB+j\rangle|P(v,j)\rangle.
\end{equation}

$\mathcal{Q}_0$ expands node identifiers to their children, while $\mathcal{Q}_1$ loads payloads. Under the classical-write/quantum-read QRAM model \citep{liu2024quantumbptree}, each call operates in $O(1)$ time. Offline index construction requires $O(N)$ storage operations following standard classical sorting.

For a query $\mathcal{W}$, the predicate oracle is:
\begin{equation}
  \mathcal{O}_{\mathcal{W}}|z_i\rangle|r_i\rangle|b\rangle =|z_i\rangle|r_i\rangle|b\oplus [z_i\in\mathcal{W}]\rangle,
\end{equation}
where the bracket evaluates to $1$ if all dimensions satisfy the predicate. Applying $\mathcal{O}_{\mathcal{W}}$ to a superposition of $m$ candidate entries and measuring the flag yields $|\Psi_{\mathcal{W}}\rangle$ with probability $k/m$. The expected post-selection time is then $O(m/k)$.

\subsection{Quantum Range Query}

Based on the aforementioned designs, QRQ executes in two steps: 1) \textbf{Classical Scan:} The index is descended to identify a candidate cover $\mathcal{C}$. Disjoint nodes are discarded, and fully contained nodes are accepted. Partially intersecting nodes are iteratively expanded until the frontier reaches the packed-index limit $q=|\mathcal{C}|$. This pruning is exact and discards no valid answers. 2.) {Quantum Expansion:} Candidate identifiers are prepared in superposition:
  \begin{equation}
    |\psi_0\rangle=\frac{1}{\sqrt{q}} \sum_{v\in\mathcal{C}}|v\rangle.
  \end{equation}

Moreover, Hadamard gates and repeated $\mathcal{Q}_0$ calls expand this state to descendant leaves. Applying $\mathcal{Q}_1$ yields:
\begin{equation}
  |\psi_{\mathrm{cand}}\rangle =\frac{1}{\sqrt{m}} \sum_{i\in\mathcal{R}(\mathcal{C})}|z_i\rangle|r_i\rangle,
\end{equation}
where $\mathcal{R}(\mathcal{C})$ includes both valid and dummy entries. Post-selecting with $\mathcal{O}_{\mathcal{W}}$ strictly isolates $\mathcal{A}(\mathcal{W})$. Because packed occupancy guarantees $k\geq\rho m$ (for a constant $\rho>0$), expected post-selection trials remain $O(1)$. 

Overall query complexity is:
\begin{equation}
  T_{\mathrm{query}}=O(h+q)+O\!\left(h\frac{m}{k}\right).
\end{equation}
%For balanced indices with constant dimensionality and bounded overlap, $h=O(\log_B N)$, simplifying the bound to $O(\log_B N)$. The traditional $O(k)$ result-size penalty is bypassed by preserving the answer in superposition. 

Herein, we implemented this QRQ with three classic tree-based indexes for spatiotemporal trajectory datasets.

\begin{itemize}
  \item \textbf{Packed R-tree:} Internal MBRs are stored in $\mathcal{Q}_0$. The classical stage halts at fully covered subtrees, while the quantum stage expands survivors and applies leaf predicates. Due to spatial overlap, empirical performance is driven by the candidate ratio $m/k$.
  \item \textbf{TB-tree:} Trajectory segments are grouped chronologically with routing bounds $g(v)=(t_{\min},t_{\max},\mathrm{MBR}_v)$. Temporal sorting heavily filters the global search, and quantum expansion simultaneously loads spatially viable segments. Payloads retain trajectory IDs and offsets to preserve temporal continuity.
  \item \textbf{KD-tree:} Disjoint recursive partitioning allows precise classical tracing. Non-overlapping regions yield a smaller $q$ and superior $m/k$ for well-balanced, low-dimensional data, though highly correlated dimensions may degrade performance to $O(h+m/k)$.
\end{itemize}

Moreover, Algorithm \ref{alg:generic-qrange} shows the pesudocode of this this implementation, where the preprocessing path follows entirely classic range query. In this context, the practical operational complexity evaluates to $O\!\left(h+q+h\frac{m}{k}\right)$, reducing to $O(\log_B N)$ for bounded frontiers and constant selectivity. 

Importantly, this acceleration is a \textbf{coherent-query speedup}, as QRAM loads target data in superposition, while geometric pruning condenses the search space from $N$ total records to a dense local cover. 

\begin{algorithm}[t]
\caption{Generic quantum range query algorithm}
\label{alg:generic-qrange}
\KwIn{Index $I$, query window $\mathcal{W}$, node oracle $\mathcal{Q}_0$, payload oracle $\mathcal{Q}_1$}
\KwOut{$|\Psi_{\mathcal{W}}\rangle=\frac{1}{\sqrt{k}}\sum_{i:z_i\in\mathcal{W}}|z_i\rangle|r_i\rangle$}
\tcp{Global classical candidate construction}
Initialize frontier $F\leftarrow\{\mathrm{root}(I)\}$ and candidate cover $\mathcal{C}\leftarrow\emptyset$\;
\While{$F\neq\emptyset$}{
  $F'\leftarrow\emptyset$\;
  \ForEach{$v\in F$}{
    \If{\textsc{Intersects}$(v,\mathcal{W})=\mathrm{false}$}{\textbf{continue}}\;
    \eIf{\textsc{Contains}$(v,\mathcal{W})=\mathrm{true}$}{
      $\mathcal{C}\leftarrow\mathcal{C}\cup\{v\}$\;
    }{
      \eIf{$v$ is a leaf}{
        $\mathcal{C}\leftarrow\mathcal{C}\cup\{v\}$\tcp*{exact leaf filtering}
      }{
        $F'\leftarrow F'\cup\mathrm{Children}(v)$\;
      }
    }
  }
  $F\leftarrow F'$\;
}
\tcp{Local quantum expansion and filtering}
Prepare $|\psi\rangle=|\mathcal{C}|^{-1/2}\sum_{v\in\mathcal{C}}|v\rangle$\;
Expand descendant slots with Hadamard gates and repeated $\mathcal{Q}_0$ calls\;
Load leaf entries with $\mathcal{Q}_1$ to obtain $|\psi_{\mathrm{cand}}\rangle$\;
Apply $\mathcal{O}_{\mathcal{W}}$ to an ancilla initialized to $|0\rangle$\;
Measure the flag and repeat preparation until the marked outcome is obtained\;
\Return the post-selected payload state\;
\end{algorithm}

\section{Results} 
\label{sec:results}

For experiment, we used three spatiotemporal trajectory datasets to test the performance of QRQ with its three variants. The datasets used are as follows: GeoLife~\cite{zheng2010geolife,zheng2008understanding} with $18{,}670$ everyday tracks ($1{,}913{,}018$ segments) representing long-extent human mobility; T-Drive~\cite{yuan2010tdrive} with $10{,}079$ Beijing taxis ($1{,}846{,}192$ segments) representing a dense, congested urban road network; and GDP Drifter~\cite{elipot2016hourly} with $555$ oceanic buoys ($1{,}208{,}891$ segments) in the North Atlantic representing sparse, basin-scale tracks. All primary evaluations benchmark at $N=1{,}048{,}576$ segments and $1\%$ target selectivity over $10{,}000$ queries.

\begin{figure}[t!]
  \centering
  \includegraphics[width=\linewidth]{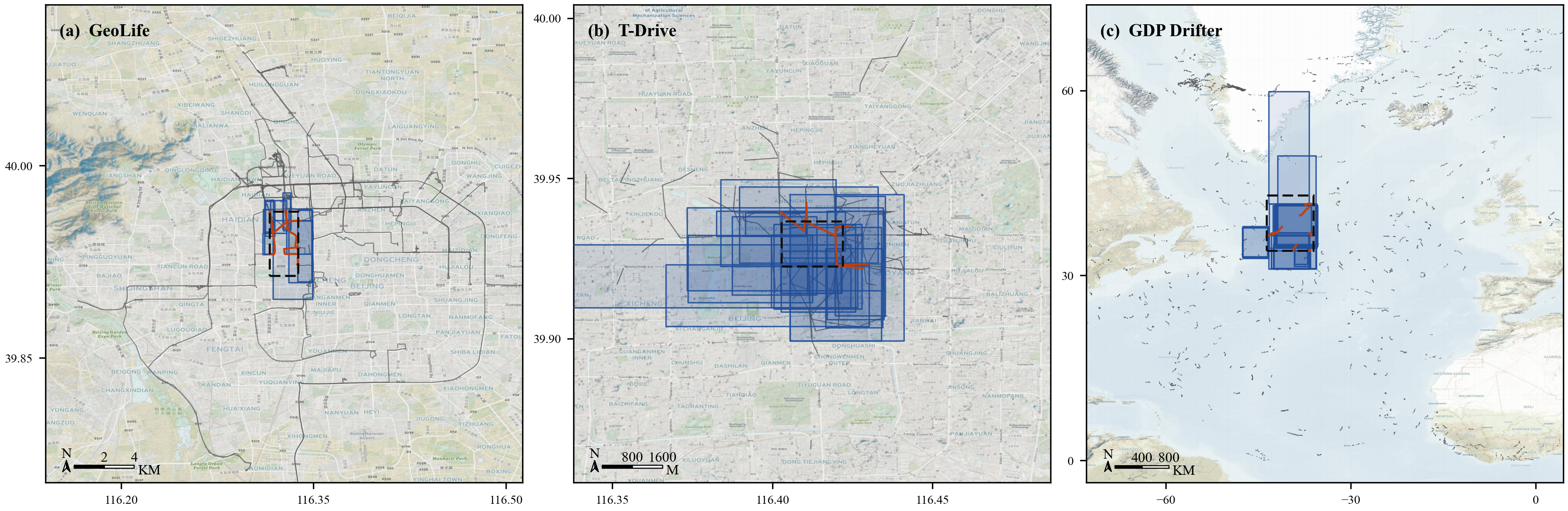}
  \vgap
  \caption{Spatial $(x,y)$ slice of a 3-D range query on the three datasets. }
  \Description{Three map panels for GeoLife, T-Drive, and GDP Drifter showing overlapping blue MBRs, a dashed query window, and orange hit segments on a terrain basemap.}
  \label{fig:retrieval}
\end{figure}

\begin{table}[t]
  \caption{QRQ speedup performances.}
  \label{tab:speedup}
  \vspace{-0.15cm}
  \small
  \begin{tabular}{llrrr}
    \toprule
    Dataset & Index & Time & Space & Space-time \\
    \midrule
    GeoLife & KD-tree & \textbf{30.35$\times$} & 4.34$\times$ & \textbf{19.85$\times$} \\
    GeoLife & R-tree & 5.21$\times$ & \textbf{4.82$\times$} & 10.36$\times$ \\
    GeoLife & TB-tree & 4.35$\times$ & 4.51$\times$ & 10.17$\times$ \\
    \addlinespace
    T-Drive & KD-tree & \textbf{8.83$\times$} & \textbf{3.02$\times$} & \textbf{9.68$\times$} \\
    T-Drive & R-tree & 2.73$\times$ & 2.03$\times$ & 3.78$\times$ \\
    T-Drive & TB-tree & 2.08$\times$ & 2.11$\times$ & 2.40$\times$ \\
    \addlinespace
    GDP Drifter & KD-tree & \textbf{8.02$\times$} & \textbf{31.72$\times$} & \textbf{64.70$\times$} \\
    GDP Drifter & R-tree & 2.35$\times$ & 16.39$\times$ & 14.10$\times$ \\
    GDP Drifter & TB-tree & 2.12$\times$ & 16.86$\times$ & 16.10$\times$ \\
    \bottomrule
  \end{tabular}\\[2pt]
  {\raggedright\footnotesize
  Note: $N=1{,}048{,}576$ segments, $1\%$ target selectivity, $10{,}000$ queries.
  Each cell is $\overline{C}/\overline{\mathcal{Q}}$.\par}
\end{table}

\begin{figure}[t!]
  \centering
  \includegraphics[width=\linewidth]{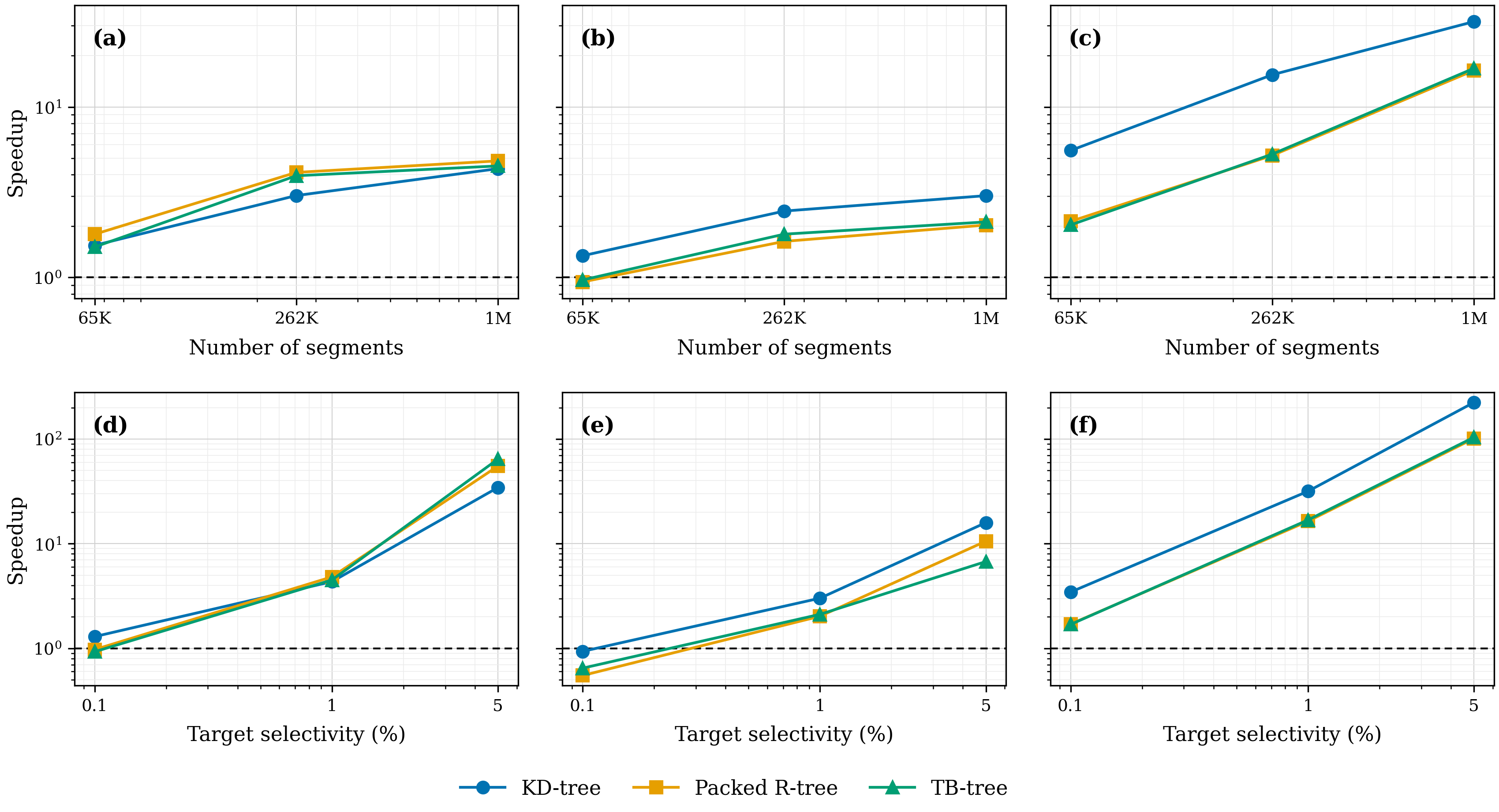}
  \vgap
  \caption{Spatial-query model speedup versus data size at $1\%$ target selectivity (a--c) and versus target selectivity (d--f). Columns are GeoLife, T-Drive, and GDP Drifter. The dashed line is $1\times$.}
  \Description{Six line charts of speedup versus number of segments and versus selectivity for KD-tree, packed R-tree, and TB-tree on three datasets, with a dashed line at speedup equal to one.}
  \label{fig:metrics}
\end{figure}

Three QRA variants achieves consistent I/O speedups ranging from $2.03\times$ to $64.70\times$ across all $27$ configurations (\Cref{tab:speedup}). Performance diverges strictly based on data geometry and spatial overlap (\Cref{fig:retrieval}). In scenarios with high overlap (GeoLife and T-Drive), dense urban trajectories create congested minimum bounding rectangles (MBRs). Queries frequently intersect boundaries only partially, which forces classical traversal to open numerous children and delays quantum post-selection, yielding modest speedups (e.g., $2.03\times$ to $4.82\times$ for space queries). Conversely, in high-sparsity scenarios (GDP Drifter), sparse oceanic tracks generate large MBRs containing minimal localized data. Queries frequently encompass these entirely, triggering containment witnesses. The quantum scan processes these fully covered subtrees in superposition without enumerating individual leaves, drastically reducing I/O and yielding peak speedups up to $64.70\times$ for space-time KD-trees.

As shown in \Cref{fig:metrics}, speedup scales positively with dataset size ($N$). However, quantum advantage requires a sufficient target result volume ($k$) to amortize post-selection overhead. At small scales ($N=65{,}000$) or low selectivities (e.g., $0.1\%$), speedups on dense datasets like T-Drive drop below $1\times$. Conversely, the high incidence of containment witnesses in the sparse drifter data preserves net I/O advantages even under strict query selectivities.

\section{Conclusion} \label{sec:conclude}

This paper presented QRQ, a hybrid quantum-classical range query for large-scale trajectory geodata. QRQ leaves the packed R-tree, TB-tree, and KD-tree unchanged and answers exactly the same segment-window queries, changing only the access path. Node slots, including padding, are read through a classical-write, quantum-read QRAM, the walk stops at the first containment or intersection witness, and the answer is returned as a superposition rather than an enumerated list. Based on three trajectory datasets (e.g., GeoLife, T-Drive, and GDP Drifter), the QRQ speedup ranges from $2.03\times$ to $64.70\times$. Moreover, a key lessons learning is that quantum advantage in spatial range query is governed by index geometry and query scale rather than by data volume alone, since containment witnesses in sparse, wide-extent data amortize post-selection quickly, whereas the overlapping MBRs of dense urban road networks delay them. Looking ahead, extending the index keys to semantic attributes would complete the spatial-temporal-semantic vision of this paper, and because QRQ keeps its answers in superposition it can feed downstream quantum analytics directly and serve as a natural retrieval primitive for queries composed by LLMs and geo-foundation models.

\begin{acks}
This work was supported by the Start-Up Grant (SUG) project “Geospatial Artificial Intelligence for Climate Resilient Urban Environment” from the National University of Singapore (E-109-00-0036-01), and the MoE Tier1 project "Assessing Urban Flood Resilience against Climate Extreme with GeoAI in Southeast Asia" (A-8004907-00-00).
\end{acks}
\hspace{-0.8cm}

\vspace{0.1cm}
\looseness=-1
% \noindent \textbf{Acknowledgement}: This work is supported by.
%%
%% The next two lines define the bibliography style to be used, and
%% the bibliography file.
\vspace{-0.3cm}
\bibliographystyle{ACM-Reference-Format}
\bibliography{reference}

\end{document}